# Momentum-space engineering in van der Waals heterostructures

Xiao Li[1,2*], Astrid Weston[1,2*], James G. McHugh[1,2*], Michael D. Thompson[3], Matthew Watson[4], Yunze Gao[1,2], Nicholas Clark[1,2], Laxman Nagi Reddy[4,5], Samuel Magorrian[5,6], Wendong Wang[1,2], Zhiyuan Wang[1,2], Sofia Konyzheva[1,2], Francisco Selles[1,2], Amy Carl[1,2], Katarzyna J. Salamon[1,2], Alex Summerfield[2], Nicholas Hine[5], Andre K. Geim[1,2], Neil R. Wilson[5], Jonathan R. Prance[3†], Andrey V. Kretinin[1,7†], Vladimir I. Fal'ko[1,2†], Roman Gorbachev[1,2†]

[1] *National Graphene Institute, University of Manchester, Oxford Road, Manchester M13 9PL, United Kingdom*

[2]*Department of Physics and Astronomy, University of Manchester, Oxford Road, Manchester M13 9PL, United Kingdom*

[3]*Department of Physics, Lancaster University, Lancaster LA1 4YB, UK*

[4]*Diamond Light Source, Division of Science, Didcot OX11 0DE, UK*

[5]*Department of Physics, University of Warwick, Coventry CV4 7AL, UK*

[6]*Hartree Centre, STFC Rutherford Appleton Laboratory, Harwell Campus, Didcot, OX11 0QX*

[7]*Department of Materials, University of Manchester, Oxford Road, Manchester M13 9PL, United Kingdom*

* These authors contributed equally to this work.

† Corresponding author, e-mails:

roman@manchester.ac.uk, vladimir.falko@manchester.ac.uk, andrey.kretinin@manchester.ac.uk, j.prance@lancaster.ac.uk

**Abstract**

Twenty years of graphene research have established Dirac fermions as a versatile platform for exploring quantum topological phenomena. Yet control over graphene's low-energy spectrum mostly follows the route of miniband engineering using real-space superlattices. Here we introduce a complementary approach based on momentum-space 'Umklapp' band engineering. This is realised by matching reciprocal space positions of band edges of two adjacent highly incommensurate 2D crystals to coincide upon a boost from their Bragg vectors, creating resonant interlayer hybridisation that qualitatively reconstructs their spectra. To demonstrate this approach experimentally, we use monolayer graphene on $MoS_2$, twisted such that $MoS_2$ Q-valley states are brought by Umklapp into energy–momentum resonance with graphene's Dirac cones. The ensuing hybridization results in the formation of a clear additional Dirac-like spectral feature, identified by a distinct Landau-level fan and Shubnikov–de Haas oscillations characteristic to Berry phase π electrons. These new topological spectral features create a spin-valley locked Ising system which is robust against small twist-angle variations that can be further tuned by an external displacement field. More broadly, Umklapp band engineering offers a practical strategy for tailoring the electronic structure in van der Waals heterostructures, expanding the design space of quantum materials beyond moiré superlattices and weak proximity effects.

## Introduction

After two decades of intense research, graphene continues to attract unrivalled community interest. Recent advances in its electronic quality place it not only ahead of many two-dimensional (2D) peers, but also on par with the best semiconductor-based heterostructures[1]. Despite this progress, tailoring graphene's celebrated linear dispersion remains challenging and, so far, has been achieved primarily by real-space band engineering through imposing an additional long-range periodicity, that is, a superlattice. The most common route is to overlay nearly commensurate 2D crystals, thereby forming a long-range moiré pattern that folds the Brillouin zone and generates minibands. For instance, the superlattice between graphene and hexagonal boron nitride (hBN) creates secondary Dirac points at miniband edges, with a complex Hofstadter spectrum emerging in high magnetic fields[2,3]. Furthermore, twisting two layers of graphene can produce flat moiré minibands that have become a fertile ground for landmark discoveries, including unconventional superconductivity, correlated insulating phases, and diverse magnetic orders, establishing "twistronics" as a vibrant field for exploring many-body physics[4–6]. These striking moiré phenomena demand close lattice matching and exceptionally fine control of the twist angle and are strongly influenced by structural relaxation and reconstruction, limiting uniformity and device-to-device reproducibility.

Tailoring of graphene's electronic properties can also be done by proximity coupling in van der Waals (vdW) heterostructures, which modifies graphene's low-energy states through interlayer proximity (tunnelling) rather than due to Bragg scattering from moiré superlattice. A prominent example is graphene coupled to transition metal dichalcogenides (TMDs), a widely studied system for research on electron transport[7,8], spintronics[9], optoelectronics[10] and related directions. In particular, placing graphene on semiconducting TMDs has been used to imprint spin-orbit coupling (SOC) onto graphene's Dirac fermions[9,11–23]. These proximity effects are usually weak, being mediated by virtual tunnelling from graphene's π-orbitals into energetically remote TMD bands. They can endow graphene's electrons with a small spin-orbit splitting but leave their low energy Dirac dispersion essentially intact[24,25].

Here, we demonstrate an alternative approach that provides a strong interlayer coupling, by bringing graphene and a TMD crystal band into energy-momentum resonance that strongly mixes and, therefore, reconstructs their states at the band energies accessible at moderate carrier densities. Rather than relying on real-space superlattices, we use twist-controlled interlayer hybridisation enabled by Umklapp-type momentum transfer. Specifically, we use 2H-$MoS_2$ films, where the conduction-band minimum lies in the Q valleys located approximately midway between the $\Gamma$ point and TMD's Brillouin-zone corners. In few-layer $MoS_2$, this Q-valley minimum can by just 100 meV away from graphene's Dirac point, enabling the resonance at gate-accessible electron densities. We then choose the relative orientation of the graphene and $MoS_2$ lattices to maximize Umklapp-assisted interlayer tunnelling for the band reconstruction, which need not be confined a narrowly defined twist angle.

## Main text

Due to the large difference between the lattice constants of the two crystals ($a_{MoS2}$ = 3.16 Å vs $a_{Gr}$ = 2.46 Å), simultaneous energy-momentum matching between graphene's K valleys and the $MoS_2$ Q valleys can only be realized in the second Brillouin zone, as illustrated in Fig. 1a by the overlaid reciprocal-space maps for $MoS_2$ and graphene at optimal twist angles $\theta \approx 18.6°$ and 41.4°. Due to the $C_3$ symmetry of 2H-$MoS_2$, graphene's K-valley (a triplet of equivalent Brillouin zone corners in Fig. 1a) couples resonantly not to one but to three inequivalent Q valleys of $MoS_2$. In terms of electrons in the first Brillouin zone of $MoS_2$, their hybridisation with Dirac states in graphene is mediated by the interlayer tunnelling accompanied by Bragg scattering from the $MoS_2$ lattice, that is, by an Umklapp process. Below, we refer to this as Umklapp hybridisation. As we show below, the resulting multiband mixing and anticrossings generate additional Dirac-like features as illustrated in Fig. 1b. By the same mechanism, reciprocal 'ghost anticrossings' are expected in the first Brillouin zone of the $MoS_2$ spectrum, analogous to those observed recently by ARPES at the valence band energies well away from the transport window relevant for electronics and spintronics[26,27]. Here, we benefit from a relatively small offset between graphene's Dirac point and the $MoS_2$ conduction band[28], so that the resonant spectral reconstruction in our case can be reached by standard electrostatic gating and therefore probed by electron transport. Only for sufficiently large misalignment does the relevant band intersection shift upward considerably (e.g., by ~50 meV for a 5° misalignment from optimal angle, See Supplementary Section 6) and become inaccessible to gating, making the resonant hybridisation regime elusive despite the same underlying band offset.

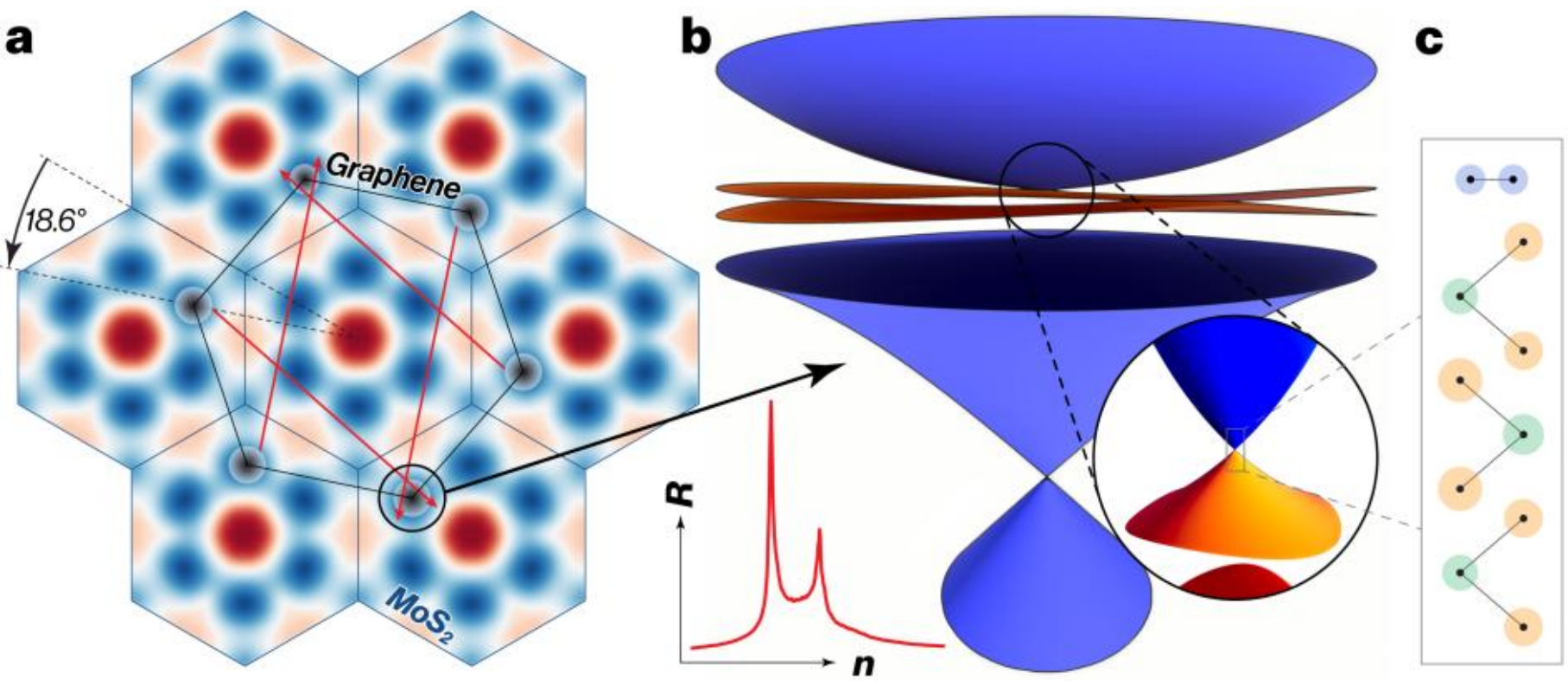

***Figure 1: Engineering of the band structure overlap in graphene/$MoS_2$ heterostructures. a**, Schematic overlay of tight-binding band structures for the lowest conduction sub-band of 3L $MoS_2$ and graphene with a 18.6° twist. Blue and black hexagons represent the $MoS_2$ and graphene BZs, respectively. Red arrows indicate resonant mixing of three $Q_{MoS2}$-valleys with each of the unique K points in graphene. **b**, A three-dimensional view of the hybridised band structure in the vicinity of the graphene K point obtained using our model, with blue denoting states primarily related to graphene and red to $MoS_2$. The inset stresses that the reconstructed spectrum is essentially a Weyl-Dirac semimetal. **c**, Here, we show the side view of the studied structure to highlight that the emergent Dirac-like feature involves electron states spread over all constituent layers of graphene and 3L $MoS_2$, with their weight on each atomic layer indicated by the area of the coloured circles.*

For the experimental observation of resonant Umklapp hybridisation, we compare electronic transport characteristics of two heterostructures with distinctly different twist angles between $MoS_2$ trilayer (3L) and monolayer graphene. To minimise sample-dependent variations, the heterostructures have been assembled from the same exfoliated crystals: one half of a large graphene layer was stacked with the 3L $MoS_2$ targeted at ≈18.6° twist angle (accuracy is limited by the optical imaging of the crystalline edges), whereas the other was picked up with an additional 30° rotation. Since the chosen crystals typically cleave along either zigzag or armchair crystallographic directions, the implemented method ensured that one of the two samples has the optimal Gr/$MoS_2$ orientation, while the other one is substantially misaligned and serves as a 'non-resonant' reference. The resulting heterostructures, encapsulated in hBN crystals on both sides, were placed over graphite back gates, and electrical contacts were created using the standard lithographic process[29], Fig.2a. An additional, third, layer of thick (~ 60 nm) hBN was transferred over both devices so that the top gates provide uniform gating of the entire Hall bar structure, avoiding artefacts that may arise in ungated graphene leads at high magnetic fields[30]. The Hall bar width was maintained at approximately 2 µm to strike a balance between homogeneity[31] and high carrier mobility of >$10^5$ $cm^2$ $V^{-1}s^{-1}$ (see Supplementary Section 1 for nanofabrication and characterisation).

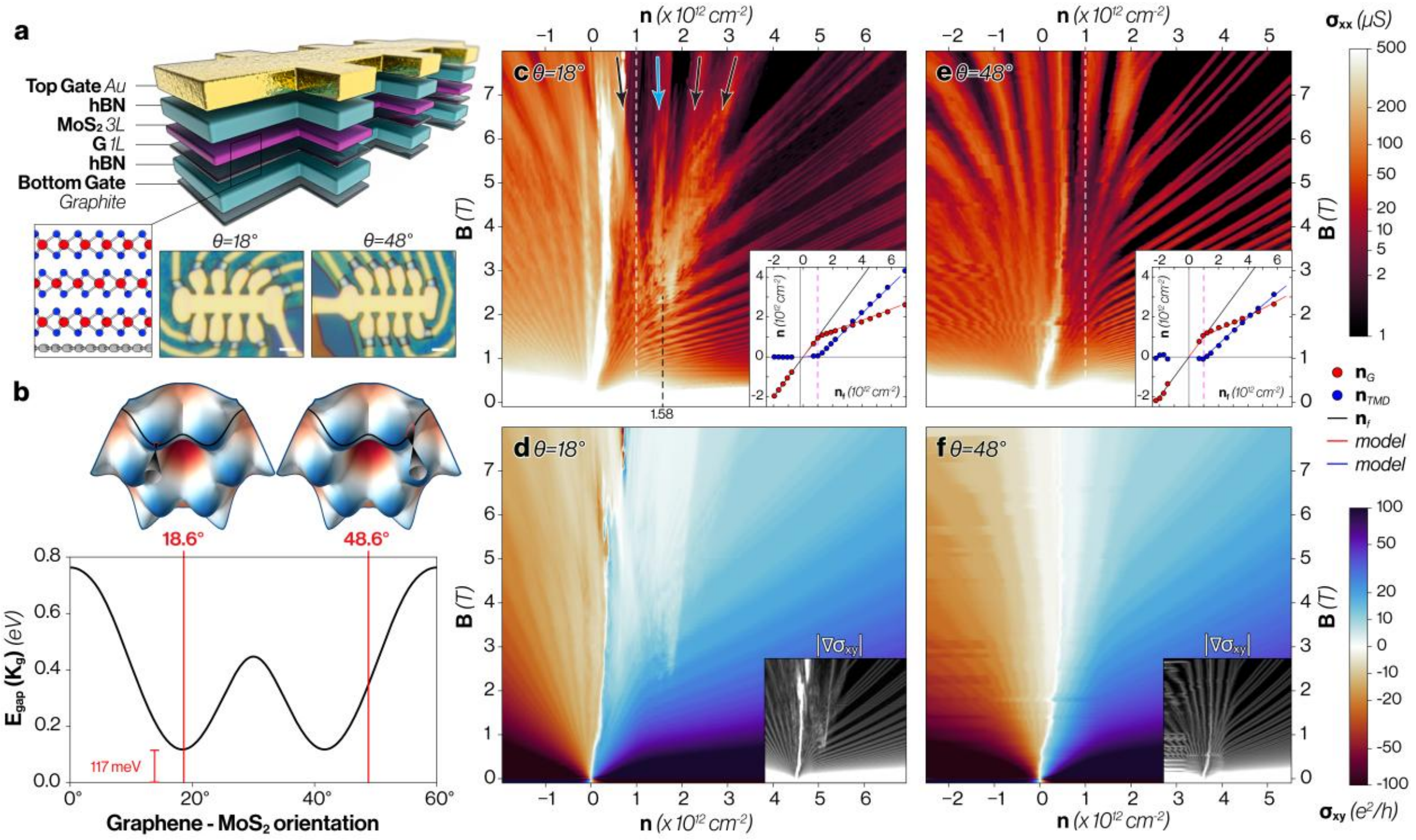

***Figure 2: Electronic transport signatures of resonant Umklapp hybridisation in 3L $MoS_2$/graphene heterostructures with twist angle of 18±0.5°. a**, Schematic representation of the studied devices (top) and their optical micrographs (bottom). The two devices have been assembled from the same crystals with different twist angles, scale bars are 2 µm. **b**, top - overlay of tight-binding result for the lowest 3L $MoS_2$ subband and graphene Dirac cones for the experimentally realized twist angles, illustrating the region in K space where the resonant hybridisation should occur. Bottom – slice of the lowest 3L $MoS_2$ subband along Q-Q direction, illustrating the energy where resonant hybridisation is expected to occur as a function of the twist angle. The "optimal" energy alignment of $\Delta E$ = 117 meV has been extracted from our transport measurements. **c**, **e**, Longitudinal conductivity ($\sigma_{xx}$) as a function of out-of-plane magnetic field (B) and carrier density (n) at zero displacement field ($D/\varepsilon_0$ = 0 V $nm^{-1}$) for the 18±0.5° device (**c**) and the misaligned reference device (**e**). Arrows indicate additional LL due to the re-entrant Dirac point. White dashed lines indicate the primary graphene LL bending position and the black dashed line denotes the zero-field extrapolation for the secondary LL originating from the re-entrant Dirac point. Insets: Carrier density analysis showing the total carrier density ($n_f$) induced by both gates, the graphene carrier density ($n_G$) extracted from the quantum oscillations, and the 3L $MoS_2$ carrier density determined as their difference ($n_{TMD} = n_f - n_G$). Pink dashed lines mark the onset of $MoS_2$ conduction band. Solid red and blue lines are the result of our self-consistent electrostatic modelling that takes into account the hybridised density of states, see Supplementary Section 8. **d**, **f**, Corresponding Hall conductivity ($\sigma_{xy}$) maps illustrating the clear presence of the second fan in the 18° sample (**d**). The insets show the derivative signal for clarity. All measurements have been performed at T = 20 mK.*

We first consider the longitudinal, $\sigma_{xx}$ and transversal, $\sigma_{xy}$ conductivities studied as a function of the total carrier density, $n$, and perpendicular magnetic field, B, in the reference device, Fig. 2e. The use of double gating allows us to cancel out the displacement field across the Gr/$MoS_2$ heterostructure (*i.e.*, working in the regime of $D/\varepsilon_0$ = 0 V nm$^{-1}$), while controlling total carrier density, $n$, which is divided between graphene and $MoS_2$, $n = n_{\mathrm{Gr}} + n_{\mathrm{MoS2}}$, where $n_{\mathrm{Gr}}$ corresponds to the carrier density in graphene and $n_{\mathrm{MoS2}}$ in 3L $MoS_2$. At low density the conventional Landau level (LL) fan diagram (Fig. 2e) can be seen, followed by a distinct bending of LLs fan at $n \sim 1\times10^{12}$ cm$^{-2}$ (corresponding to $E_F \approx 117$ meV). At this point, the Fermi level reaches the lowest conduction sub-band in $MoS_2$, so that a sudden decrease in the $dn_G/dn$ indicates a substantial portion of the added electrons now enter the $MoS_2$. However, due to the much higher mobility of electrons in graphene -- as compared to the heavy carriers in $MoS_2$ -- the overall conductivity and the Shubnikov-de Haas oscillations are dominated by the graphene layer. At higher electron densities ($n > 3.4 \times 10^{12}$ cm$^{-2}$), a clear four-fold sub-LL splitting emerges, indicative of lifted spin and valley degeneracies. This dramatic change in the graphene LL resolution is likely due to neutralising charged impurities in $MoS_2$ upon filling its conduction band with electrons.

In a stark contrast, the device with the optimally oriented Gr/$MoS_2$ reveals new distinct transport features, slightly above the expected $Q_{\mathrm{MoS2}}$-valley band edge of 3L $MoS_2$, Fig. 2c. Those consist of an additional, clearly resolved fan of LLs with both positive and negative slopes (reflecting filling factors of such LLs). This demonstrates that both electron- and hole-like states are present near $n \approx 1.6\times10^{12}$ cm$^{-2}$, well above the primary Dirac point of graphene, where, normally, only electron-like states in both $MoS_2$ and graphene are expected. This behaviour points towards a new topological band feature emerging in the semimetallic regime. Furthermore, the high visibility of the additional LL fan indicates that the corresponding LLs include substantial contributions from highly mobile graphene states, since the pure $MoS_2$ states do not noticeably contribute to the conductivity of the reference sample. Judging by the vertical feature observed in $\sigma_{xx}$ at $n \approx 1.6\times10^{12}$ cm$^{-2}$, which is reminiscent of the zeroth LL in graphene, we relate that LL fan to the Dirac-like point produced by the resonant Umklapp hybridisation. For reproducibility measurements in other devices (both aligned and non-aligned) see Supplementary Section 2. The coexisting electron- and hole-like carriers in the optimally aligned heterostructure are further verified by the consistent behaviour of Hall conductivity ($\sigma_{xy}$), apparent in Fig. 2d. It occurs between filling factors 0 and +2 traced from the primary Dirac point in graphene, where normally only electronic states are expected. However, when the negative LL from the secondary fan intersects this region at B ≈ 6T, a clear pocket of negative charge is observed in Fig. 2d, confirming that the new spectral feature is strong enough to dominate the Hall signal over the primary Dirac point. The aforementioned features in conductivity can be considerably enhanced if the carrier mobility of graphene is suppressed by making the conducting channel narrower, thereby highlighting the contribution from $MoS_2$ electrons. Thus, in our results measured on 0.6 µm-wide device, the re-entrant Dirac point dominates over the primary one in both zero and high magnetic fields, see Supplementary Fig. S6.

For the quantitative description of the anticrossings produced by the resonant Umklapp hybridisation of graphene $K_{\mathrm{Gr}}$ and TMD $Q_{\mathrm{MoS2}}$ valleys, in particular, those specific for 3L-$MoS_2$, we employ the following 'minimal' model. This model combines a k·p theory description of the lowest subband in an N-layer 2H-$MoS_2$ (NL-$MoS_2$) film with interlayer Umklapp hybridisation between graphene and $MoS_2$. For each of graphene's two K-valleys ($\tau$K, with valley index $\tau$=±), three inequivalent $\mathrm{Q}_{\mathrm{MoS2}}^{\mathrm{j=1,2,3}}$ valleys are folded according to their respective Umklapp conditions,

$$\vec{k} + \vec{G}_{Gr} = \vec{\kappa}_j + \vec{G}_{MoS2}, \quad (1)$$

where $\vec{k}$ is the electron wave vector in graphene and $\vec{\kappa}_j$ in NL-$MoS_2$ (chosen near the bottoms of three $Q_{MoS2}^{j=1,2,3}$ valleys); $\vec{G}_{Gr}$ and $\vec{G}_{MoS2}$ are reciprocal lattice vectors linked to the orientations of graphene and $MoS_2$ lattices. As interlayer tunnelling decays rapidly with the in-plane wave number, we need only account for the shortest reciprocal lattice vectors, resulting in the following (3N+2)× (3N+2) Hamiltonian matrix:

$$\mathcal{H}_{NL,s}^{\tau}(\vec{k}) = \begin{pmatrix} H_{N,\theta}^{s\tau}(\vec{\kappa}_j) & T_N^{s\tau\dagger} \\ T_N^{s\tau} & H_{Gr}^{\tau K}(\vec{k}) \end{pmatrix}, \; T_N^{s\tau} = \begin{pmatrix} \ddots & \vdots & \vdots & \vdots \\ \cdots & t_0 & t_0 & t_0 \\ \cdots & t_0 & e^{-\frac{i2\pi}{3}} t_0 & e^{-\frac{i4\pi}{3}} t_0 \end{pmatrix}. \quad (2)$$

Here, $H_{Gr}^{\tau}(\boldsymbol{k}) = \hbar v(\tau k_x \sigma_x + k_y \sigma_y)$ takes into account sublattice composition of the electron states in graphene, and a block-diagonal 3N×3N matrix, $H_{N,\theta}^{s\tau}(\vec{\kappa}_j)$, describes $MoS_2$ band edge states: it accounts for the effective mass anisotropy ( $m_{zigzag} \approx 0.66\, m_e, 0.61\, m_e$ and $m_{armchair} \approx 1.31\, m_e, 1.21 m_e$ of lower and upper spin-split edges) rotated by $120^{o} \times j$ in $Q_{MoS2}^{j=1,2,3}$ valleys; a substantial spin-orbit splitting, $\Delta_{SOC}$= 68 meV, of monolayer Q-point band edges which, individually, lack inversion symmetry; and the hybridisation of $Q_{MoS2}^{j=1,2,3}$ valley states between consecutive monolayers. Finally, an interlayer tunnelling block, $T$, accounts for the interaction between graphene and the closest layer of the $MoS_2$, where $t$ is the Umklapp tunnelling parameter, and all omitted entries are zero. The strength of the tunnelling parameter was parametrised with the help of large-scale density functional theory (DFT) calculations implemented in the ONETEP code (Supplementary Section 3, 4, 5 for additional DFT fitting and model details). By comparing band anticrossings in Fig. S14 with the model in Eq. (2), for each spin projection, we determined the value of the mixing parameter, $t$~25 meV, which we use to discuss the experimentally detected features of the optimally aligned Gr/3L $MoS_2$ heterostructure ($\theta \approx 18^{o}$).

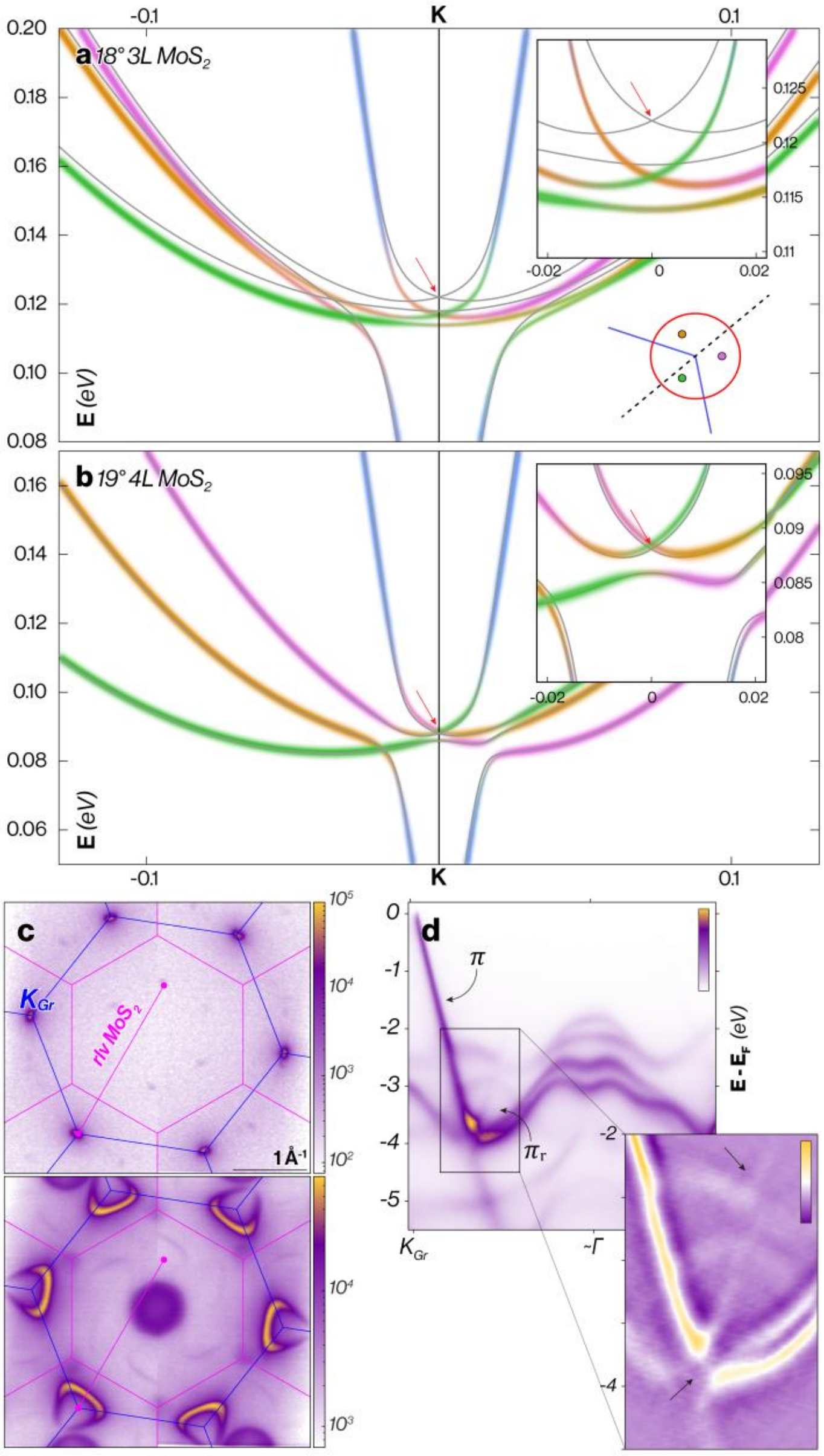

***Figure 3**: **Anticrossing of graphene and $MoS_2$ states in an aligned heterostructures**. **a**, Spin-resolved anticrossing between K graphene and 3L-$MoS_2$ Q valley for $\theta = 18^o$, and **b**, 4L-$MoS_2$ Q valley for $\theta = 19^o$, computed using the minimal model in Eq. (2) with t = 25 meV. Bands are coloured by their projection onto reference Bloch states: blue (graphene), orange ($MoS_2$ direct Q-valley edge, $\vec{G}_{MoS2} = 0$), and green/purple (folded replicas, $\vec{G}_{MoS2} \neq 0$). The grey bands show dispersion for the opposite spin, with valley colouring omitted for simplicity. Red arrows mark the re-entrant Dirac point. The top insets in a, b show magnified views of the re-entrant Dirac point. The band path is shown in the bottom insets, with a reference radius of 0.06 $Å^{-1}$ shown in red. **c**, μARPES constant energy maps of Gr/3L $MoS_2$, with a twist angle of 21.5 $^o$, at $E_F$ (upper), and at $E$-$E_F$ = -1.65 eV (lower), which corresponds to near the top of the valence band at the $K_{MoS2}$ valleys. The backfolding of Gr bands due Umklapp scattering by $MoS_2$ can be seen in the $1^{st}$ BZ of $MoS_2$. The data at $k_x$<0 are symmetrized by twofold rotation. **d**, Dispersion aligned to the Γ-$K_{Gr}$ direction showing the graphene π band crossing the $MoS_2$ valence bands and an Umklapp scattered replica graphene band, $\pi_r$. Inset lower right is a curvature analysis, highlighting multiple hybridisation features where the bands of Gr and $MoS_2$ anticross in the occupied valence bands. All ARPES data were measured with a photon energy of 140 eV.*

Compared to 2H-$MoS_2$ bilayers, trilayers have a slightly lower energy of the $\mathrm{Q}_{\mathrm{MoS2}}^{\mathrm{j=1,2,3}}$ valleys, and, because they lack inversion symmetry, the Q band edges are spin-split (Supplementary Section 5). Consequently, the offset between graphene's Dirac point, and, hence, the conditions for Umklapp resonance of spin $s = \downarrow$ and $\uparrow$ states are distinct. The resulting multiband spectrum, obtained using Eq. (2), is shown in Fig. 3a, where the computed dispersions near $K_{Gr}$ are plotted along the zigzag direction. Colours of the dispersion lines in this plot indicate the composition of hybridised states for the low-energy spin orientation, with orange, purple and green denoting each of the $Q_{MoS2}^{j=1,2,3}$ valleys, while blue corresponds to graphene. The branches of bare graphene's linear dispersion approaching from below can be seen to experience the anticrossing, producing a 're-entrant' Dirac point with a characteristic velocity of $v' \approx 50{,}000$ ms$^{-1}$. Notably, the two spin, $s = \downarrow$ and $\uparrow$ bands are separated by $\delta \sim 7$ meV, roughly a fivefold enhancement over achievable proximity-induced SOC[17,21] (see Supplementary Sections 6, 7, 9 for a detailed twist-angle dependence analysis). Going up to 4L $MoS_2$/graphene heterostructures, Fig.3b, we see the inversion symmetry restored and hence the spin splitting becomes smaller, $\sim 1$ meV. The states forming this feature carry substantial weights of graphene-based orbitals, allowing them to produce a detectable contribution to conductance in our experiments. We relate them to the additional LL fan in Fig. 2c, due to the 'inverted' (hole-like) dispersion they have below the re-entrant Dirac point that covers a wave number range about 0.02 Å$^{-1}$.

To confirm the validity of our model, we employ the obtained density of states to describe the charge distribution between graphene and $MoS_2$ using self-consistent electrostatic calculations (for details see Supplementary Section 8). The result agrees well with the experimental carrier densities for graphene, $n_G$ obtained from the Shubnikov–de Haas effect and $MoS_2$, found as the difference between the capacitive model and graphene's density $n_{TMD} = n - n_G$, see insets of Fig. 2c,e. We used only one fitting parameter, the interlayer coupling of 25 meV, to obtain a good match, confirming the validity of our DFT calculations (Supplementary Section 3). Moreover, our model allows us to compute the carrier density at which the re-entrant Dirac point should emerge as $n \approx 1.7\times10^{12}$ cm$^{-2}$, in excellent agreement with our experimentally observed value of $\approx 1.58\times10^{12}$ cm$^{-2}$, Fig.2c.

Angle-resolved photoemission spectroscopy (ARPES) provides direct evidence for the Umklapp ($MoS_2$ reciprocal lattice vector) replicas of graphene Dirac cones as well as resonant interlayer Umklapp hybridization appearing as 'ghost' anti-crossings. The data presented in Fig. 3c were acquired from a Gr/3L $MoS_2$ heterostructure with the twist angle of around 21.5°. The Fermi-surface map in Fig. 3c, at the Fermi energy $E_F$, shows not only strong photoemission intensity at the six $K_{Gr}$ valleys, as expected from the graphene Dirac points, but also reveals weak intensity due to Umklapp scattered replica graphene Dirac points at positions roughly coincident with the $Q_{MoS2}$ valleys. In the second constant energy map, at an energy chosen such that intensity can be seen from the valence band edges at the $K_{MoS2}$ valleys, the $MoS_2$ valence band edge and the relative orientation of the lattices can be readily discerned. In this map, from the orientation of the dark corridors[32] in the replica graphene bands, it is clear that they are formed by scattering the $K_{Gr}$ valleys by $\vec{G}_{MoS2}$. Plotting the dispersion along the Γ-$K_{Gr}$ direction shows the graphene band extending to $E_F$ at $K_{Gr}$, as well as the characteristic valence bands of $MoS_2$, Fig. 3d. These bands are not simply overlaid, interactions between them are clear. The corresponding curvature plot[33] emphasises the anti-crossing between the graphene π band and $MoS_2$ valence band[26], and further reveals interactions between the Umklapp-scattered replica graphene $\pi_r$ band and the upper $MoS_2$ valence band, resulting in additional anti-crossings (highlighted by black arrows in Fig. 3d). This interaction between replica and primary bands confirms that the replicas originate from initial state effects rather than from final state artefacts due to scattering in the photoemission process[34], hence, are due to the resonant Umklapp hybridisation between the graphene and $MoS_2$.

To shed more light on the nature of the observed LL features in the magnetotransport measurements, we investigate another device with 4L $MoS_2$ twisted 19.5 degrees to graphene and higher electronic quality. In this device, the Hall bar width was decreased to 0.6 µm, resulting in suppression of graphene's mean free path and therefore a decrease in conductivity of the native graphene states, making the contribution from $MoS_2$-hybridised states to the overall conductivity more prominent. As a result, we observe the fan diagram where the signature of hybrid fermions can be traced all the way to low magnetic fields, Fig.4b, with a clear resistance peak at B=0, Fig. 4i.

Application of a displacement field sheds more light on the nature of the observed LL features. Since the 4L $MoS_2$ has finite thickness and its states are distributed throughout all four layers, it is possible to displace its electronic density towards one of the outer layers using an external electric field[35]. This influences the wave-function overlap between the lowest $MoS_2$ layer and graphene, providing a way of tuning the tunnelling parameter $t$ in Eq. (2). This effect can be clearly seen in Fig. 4c, as a positive displacement field, $D/\varepsilon_0$ = 0.15 V $nm^{-1}$, leads to a more pronounced re-entrant LL sequence: this is because the field pushes the band edge state wave functions towards graphene, thus, enhancing hybridisation and favouring stronger re-entrant Dirac cone features. Consequently, negative $D/\varepsilon_0$ = -0.15 V $nm^{-1}$, weakens the hybridisation and results in less defined re-entrant features in Fig. 4a, (note that secondary fan features are not visible in non-aligned devices regardless of the D-field, see SI Figs. S11 and S12). In addition, the presence of the displacement field leads to a change in the band alignment, as evidenced by the carrier density required to reach the edge of the $MoS_2$ conduction band. This means that not only the hybridisation strength, but also the anti-crossing energy can be tuned with respect to the Dirac point position.

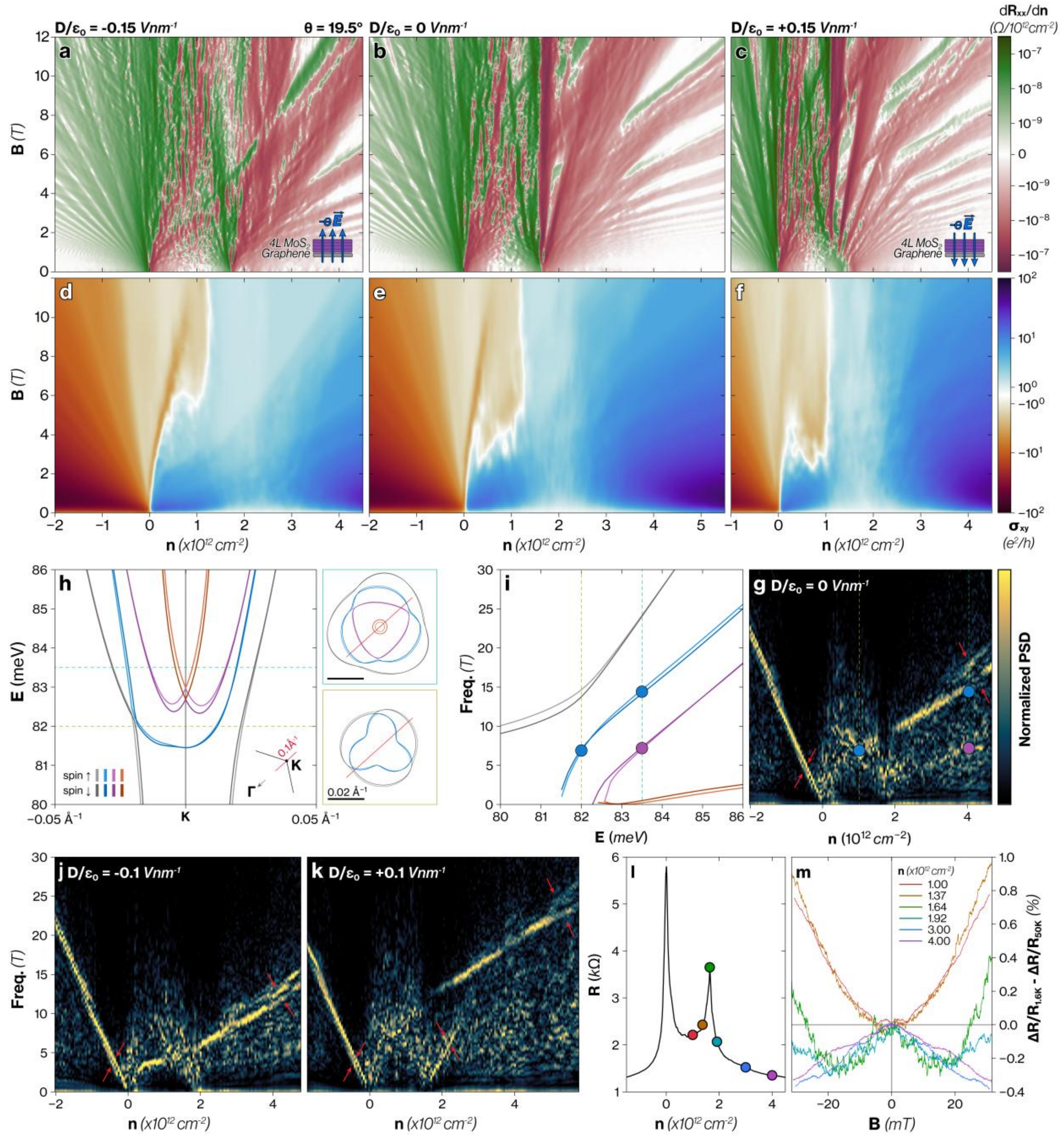

***Figure 4. Displacement field-induced changes in band hybridisation of the aligned 4L $MoS_2$/graphene heterostructure. a-c**, Derivative of longitudinal resistivity ($dR_{xx}/dn$) as a function of perpendicular magnetic field and carrier density at displacement fields $D/\varepsilon_0$ = -0.15 V nm$^{-1}$ (**a**), 0 V nm$^{-1}$ (**b**), and 0.15 V nm$^{-1}$ (**c**). **d-f**, Corresponding Hall conductivity ($\sigma_{xx}$) maps for fields $D/\varepsilon_0$ = -0.15 V nm$^{-1}$ (**d**), 0 V nm$^{-1}$ (**e**), and 0.15 V nm$^{-1}$ (**f**). **h**, Left - the bandstructure calculated for 4L $MoS_2$/graphene heterostructure, with colour identifying different subbands. Right – two constant energy slices showing the Fermi surfaces at the energies marked with dashed lines. **i**, Shubnikov–de Haas (SdH) oscillation frequencies calculated using the bandstructure shown in **h**. **g-k**, Experimental SdH frequencies for transverse fields $D/\varepsilon_0$ = 0 (**g**), -0.1 V nm$^{-1}$ (**j**), 0.1 V nm$^{-1}$ (**k**), demonstrating the presence of distinct dispersion branches of the re-entrant Dirac point. Arrows indicate a noticeable doubling of the frequencies. **l**, **m**, Magnetoresistance in a small B field demonstrates a transition between weak localisation and weak anti-localisation, as the Fermi level passes through the re-entrant Dirac point. All experimental results were acquired at T=1.6K.*

The high quality of this 4L $MoS_2$/graphene device has also enabled us to access the Shubnikov–de Haas oscillations (SdH) across the entire density range. The extracted frequency $\nu$ plotted in Fig.4g describes how the Fermi surface area, $A_F$, evolves with carrier density according to the Onsager relation $\nu = \hbar A_F/2\pi e$ (for details of the analysis see SI Figure S9). At the origin, the classical V-shape of the Dirac fermions is seen, with a similar feature also emerging at $n \approx 1.6\times10^{12}$ $cm^{-2}$ corresponding to the re-entrant Dirac point and two clear branches above > $2\times10^{12}$ $cm^{-2}$. To identify these features, we refer to our model shown in Fig.4h and the corresponding SdH frequencies calculated for each individual band, Fig.4i. Four clearly identifiable branches emerge: the grey and blue are graphene and TMD branches that undergo the anti-crossing and exchange their orbital characteristics, while the purple and orange are the outer and inner envelopes of the re-entrant Dirac point, respectively. Above the re-entrant Dirac point (at the energy marked with the dashed line), two frequencies labelled with blue and purple dots in Fig.4i closely match those observed in our experiment, Fig.4g. The two other branches are absent in the experiment: the orange branch possesses frequency ~1T is inaccessible in SdH measurement, while the grey at high energies becomes a pure $MoS_2$ branch that does not yield clear quantum oscillations. Consistently with our previous observations, the application of the negative transverse electric field noticeably suppresses the re-entrant spectral features, Fig.4j, while application of the positive field enhances them so that the second branch emerging from the re-entrant DP can be clearly seen Fig.4k. Moreover, we can resolve a noticeable twinning of the native and re-entrant Dirac branches (marked with red arrows), which agrees with our expectations regarding spin-splitting for the Gr/4L $MoS_2$ structure. Note that we observe such splitting only in structures with 4L $MoS_2$ layers, as a large spin splitting in 3L $MoS_2$ pushes the inverted-spin state outside of the accessible carrier density range. All the hybridised spectral features as well as the line splitting are absent in non-aligned devices, see Supplementary Figure S13.

Finally, our measurements of the small field magnetoresistance (Fig.4l,m) display a clear transition from weak anti-localisation (WAL) to weak localisation (WL) upon crossing of the re-entrant Dirac point. The presence of WAL in a weakly interacting regime has been previously linked to the presence of the Rashba-type spin-orbit interaction[11,36,37]. While further analysis of magnetoresistance in the presence of multiple carrier types will require a separate in-depth investigation, we nevertheless believe there is a strong indication of a substantial spin-orbit splitting in this system when its Fermi energy is tuned towards the resonant hybridisation with $MoS_2$ band edges.

**Conclusion and outlook.**

We have introduced Umklapp band engineering as a practical route to access strong interlayer hybridization in vdW heterostructures under conditions accessible by electrical measurements. This opens a complementary (to moiré superlattices) pathway towards intelligent design of band-structure singularities in lattice-mismatched stacks, by exploiting coupling between first- and higher-Brillouin-zone features. Applied to graphene and $MoS_2$, this approach produces a Weyl/Dirac-type semimetal with its distinct topological properties out of otherwise featureless electronic dispersion of the constituent crystal. The system is highly tuneable: as the carriers near the main Dirac point remain relatively unaffected, one can switch between the conventional and hybridised Dirac states with electrostatic gating and tune the hybridisation strength using the transverse electric field. While $MoS_2$ was selected here due to its favourable band alignment with graphene[28], the same principle can be applied to other TMD crystals and 2D materials in general: for instance, bilayer and trilayer graphene systems where rich many-body physics is still actively being discovered (e.g.[38,39]), imbuing those with new degrees of freedom in design of quantum properties. The proposed approach offers a good deal of flexibility: the energy scale and character of the hybridized states can be additionally engineered through strain, via chemical doping or alloying, or the layer number.

This approach also opens multiple avenues to leverage powerful TMD optical properties and excellent graphene electronic quality to design materials for novel devices. We can envisage creating conditions for coherent information exchange between $\Gamma Q$ excitons and confined electrons in bilayer graphene[40]. The hybridised branches display a substantial spin splitting that depends on the parity of the TMD layer count, resulting in spin-valley locked Ising system, which can allow for exploring "valleytronics" approaches in the THz range and electrostatically tuneable spintronic devices. The central message is that reciprocal-space matching can be used as a design knob to bring otherwise weak proximity interactions into a regime where they produce charge- and spin-transport-relevant reconstruction of graphene's electronic structure. Looking ahead, Umklapp engineering provides a route to tailor dispersions and Fermi-surface topology, and to pursue proximity-enabled functionalities (such as enhanced spin-orbit response and topological band features) in reproducible, gate-configurable device architectures.

## Methods

### Device Fabrication

The devices with dual-gated structure have been assembled as follows. Monolayer graphene and few-layer $MoS_2$ crystals were exfoliated on PMMA/Si wafers. To ensure precise angular control during stacking, we exclusively selected flakes with multiple prominent straight edges aligned with the crystal axes (indicated by internal angles of multiples of 30°, see Supplementary Section 1). The stacks were made using propylene carbonate (PC) film on polydimethylsiloxane (PDMS) stamp to first pick up the top hBN and $MoS_2$ crystals. Then, the long and straight edge of the first graphene piece was aligned with the edge of $MoS_2$ at 18.6°twist angle. The source substrate was then further rotated by 30°, and a second piece of graphene was picked up avoiding an overlap with the first piece. Then, the bottom hBN was picked up and the entire stack was released on either a graphite bottom gate or $SiO_2$/Si wafer (using Si at the global back-gate). Following electron beam lithography and dry etching with $CHF_3/O_2$ plasma the edge contacts (Cr/Au, 3 nm/55 nm) were deposited. The active channel and contacts were then fully encapsulated by transferring an additional top hBN layer ($\sim$ 60 nm thickness). Finally, a top gate (Cr/Au, 1 nm/50 nm) was fabricated on top of the encapsulation layer, and the hall bar geometry was defined with a dry etch in $CHF_3/O_2$ plasma. For more fabrication details and ARPES device fabrication see Supplementary Section 1.

### Measurements

Magnetotransport measurements were carried out using standard low-frequency lock-in amplifier techniques. Top-gate voltage ($V_t$) and bottom-gate voltage ($V_b$) are swept to adjust carrier density $n = (C_b V_b + C_t V_t)/e + n_0$ and displacement field $D/\varepsilon_0 = (C_b V_b - C_t V_t)/2\varepsilon_0$, where $C_t$ and $C_b$ are top- and bottom-gate capacitance per area.

### µARPES

Micro-ARPES measurements were performed on the nano branch of the I05 beamline, using the capillary mirror optic to focus the beam to a size of 4x4 um FWHM. The combined energy resolution for the data shown in Fig. 3 was ~50 meV and the beam polarisation was set to linear horizontal. The sample temperature was ~30 K.

**Data Availability:** Raw images and underlying data will be made available to download upon publication.

**Acknowledgements:** We acknowledge support from European Union's Horizon 2020 Research and Innovation programme: European Graphene Flagship EC-FET Core3 Project (grant agreement no. 881603 R.G., A.K. and V.F.), European Quantum Flagship Project 2DSIPC (820378, R.G. and V.F.), ERC Consolidator QTWIST (no. 101001515, R.G.), and the Royal Society (R.G). In addition, we acknowledge support from EPSRC (grant numbers EP/V007033/1 R.G. and V.F., EP/V036343/1 R.G. and A.K., EP/S030719/1 R.G. and V.F., EP/T027207/1 N.R.W., EP/V000136/1 S.J.M. and N.H.), the Lloyd Register Foundation Nanotechnology Grant, CDT Graphene-NOWNANO A.C., Y.G., and China Scholarship Council (CSC) under grant numbers 201806280036 W.W, and 201908890023 Y.G. This project was supported by the Henry Royce Institute for Advanced Materials, funded through EPSRC grants EP/R00661X/1, EP/S019367/1, EP/P025021/1 and EP/P025498/1. Computing facilities were provided by the Scientific Computing Research Technology Platform of the University of Warwick through the use of the High-Performance Computing (HPC) cluster Avon, and the UK national high performance computing service, ARCHER2, for which access was obtained via the UKCP consortium and funded by EPSRC Grant ref EP/X035891/1.

**Author contributions:** R.G. and V. F. conceived the study. A.W., X.L. and W.W. produced experimental samples with help from F.S., A.C., K.S. and A.S.; X.L., M.T., Y.G., S.K., Z.W. and J.P. performed electrical measurements with help from A.K.; M.W., L.N.R. and N.W. performed ARPES measurements and data analysis. J.G.M., S.M., N.H. and V.F. conducted theoretical modelling; R.G., V.F., X.L., N.C. and A.K. wrote the manuscript. All authors contributed to the discussions and commented on the manuscript. X.L., A.W. and J.G.E. have contributed to this work equally and can claim their leading role for job applications, interviews, and talks.

**Competing interests:** The authors declare no competing interests.

**Additional Information**: Supplementary Information is available for this paper.

1. Domaretskiy, D. *et al.* Proximity screening greatly enhances electronic quality of graphene. *Nature* **644**, 646–651 (2025).
2. Dean, C. R. *et al.* Hofstadter's butterfly and the fractal quantum Hall effect in moiré superlattices. *Nature* **497**, 598–602 (2013).
3. Ponomarenko, L. A. A. *et al.* Cloning of Dirac fermions in graphene superlattices. *Nature* **497**, 594–597 (2013).
4. Cao, Y. *et al.* Unconventional superconductivity in magic-angle graphene superlattices. *Nature* **556**, 43–50 (2018).
5. Cao, Y. *et al.* Correlated insulator behaviour at half-filling in magic-angle graphene superlattices. *Nature* **556**, 80–84 (2018).
6. Bistritzer, R. & MacDonald, A. H. Moiré bands in twisted double-layer graphene. *Proc. Natl. Acad. Sci. U. S. A.* **108**, 12233–12237 (2011).
7. Banszerus, L. *et al.* Extraordinary high room-temperature carrier mobility in graphene-WSe$_2$ heterostructures. Preprint at https://doi.org/10.48550/arXiv.1909.09523 (2019).
8. Zhang, Y. *et al.* Enhanced superconductivity in spin–orbit proximitized bilayer graphene. *Nature* **613**, 268–273 (2023).
9. Garcia, J. H., Vila, M., Cummings, A. W. & Roche, S. Spin transport in graphene/transition metal dichalcogenide heterostructures. *Chem. Soc. Rev.* **47**, 3359–3379 (2018).
10. Britnell, L. *et al.* Strong light-matter interactions in heterostructures of atomically thin films. *Science* **340**, 1311–1314 (2013).
11. Wang, Z. *et al.* Strong interface-induced spin-orbit interaction in graphene on WS 2. *Nat. Commun.* **6**, (2015).
12. Avsar, A. *et al.* Spin–orbit proximity effect in graphene. *Nat. Commun.* **5**, 4875 (2014).
13. Gmitra, M. Graphene on transition-metal dichalcogenides: A platform for proximity spin-orbit physics and optospintronics. *Phys. Rev. B* **92**, (2015).

14. Wakamura, T. Strong Anisotropic Spin-Orbit Interaction Induced in Graphene by Monolayer WS2. *Phys. Rev. Lett.* **120**, (2018).

15. David, A., Rakyta, P., Kormányos, A. & Burkard, G. Induced spin-orbit coupling in twisted graphene--transition metal dichalcogenide heterobilayers: Twistronics meets spintronics. *Phys. Rev. B* **100**, 085412 (2019).

16. Island, J. O. *et al.* Spin–orbit-driven band inversion in bilayer graphene by the van der Waals proximity effect. *Nature* **571**, 85–89 (2019).

17. Masseroni, M. *et al.* Spin-orbit proximity in MoS2/bilayer graphene heterostructures. *Nat. Commun.* **15**, 9251 (2024).

18. Gerber, J. D. *et al.* Tunable Spin–Orbit Splitting in Bilayer Graphene/WSe2 Quantum Devices. *Nano Lett.* **25**, 12480–12486 (2025).

19. Icking, E. *et al.* Weak localization as probe of spin-orbit-induced spin-split bands in bilayer graphene proximity coupled to WSe$_2$. Preprint at https://doi.org/10.48550/arXiv.2505.24632 (2025).

20. Bisswanger, T., Schmidt, A., Volmer, F., Stampfer, C. & Beschoten, B. Room-temperature spin-lifetime anisotropy exceeding 60 in bilayer graphene spin valves proximity coupled to WSe$_2$. Preprint at https://doi.org/10.48550/arXiv.2505.18990 (2025).

21. Li, Y. Twist-angle dependence of the proximity spin-orbit coupling in graphene on transition-metal dichalcogenides. *Phys. Rev. B* **99**, (2019).

22. Naimer, T., Zollner, K., Gmitra, M. & Fabian, J. Twist-angle dependent proximity induced spin-orbit coupling in graphene/transition metal dichalcogenide heterostructures. *Phys. Rev. B* **104**, 195156 (2021).

23. Faria Junior, P. E. *et al.* Proximity-enhanced valley Zeeman splitting at the WS2/graphene interface. *2D Mater.* **10**, 034002 (2023).

24. Coy Diaz, H. *et al.* Direct Observation of Interlayer Hybridization and Dirac Relativistic Carriers in Graphene/MoS2 van der Waals Heterostructures. *Nano Lett.* **15**, 1135–1140 (2015).

25. Pierucci, D. *et al.* Band Alignment and Minigaps in Monolayer MoS2-Graphene van der Waals Heterostructures. *Nano Lett.* **16**, 4054–4061 (2016).

26. Magorrian, S. J. *et al.* Band alignment and interlayer hybridisation in transition metal dichalcogenide/hexagonal boron nitride heterostructures. *2D Mater.* **9**, 045036 (2022).

27. Graham, A. J. *et al.* Ghost anti-crossings caused by interlayer umklapp hybridization of bands in 2D heterostructures. *2D Mater.* **8**, 015016 (2020).

28. Gao, Y. *et al.* Tunnel junctions based on interfacial two dimensional ferroelectrics. *Nat. Commun.* **15**, 4449 (2024).

29. Wang, L. *et al.* One-dimensional electrical contact to a two-dimensional material. *Science* **342**, 614–617 (2013).

30. Williams, J. R., DiCarlo, L. & Marcus, C. M. Quantum Hall Effect in a Gate-Controlled p-n Junction of Graphene. *Science* **317**, 638–641 (2007).

31. Uri, A. *et al.* Mapping the twist-angle disorder and Landau levels in magic-angle graphene. *Nature* **581**, 47–52 (2020).

32. Mucha-Kruczyński, M. Characterization of graphene through anisotropy of constant-energy maps in angle-resolved photoemission. *Phys. Rev. B* **77**, (2008).

33. Zhang, P. *et al.* A precise method for visualizing dispersive features in image plots. *Rev. Sci. Instrum.* **82**, 043712 (2011).

34. Graham, A. J. *et al.* Conduction Band Replicas in a 2D Moiré Semiconductor Heterobilayer. *Nano Lett.* **24**, 5117–5124 (2024).

35. Zultak, J. *et al.* Ultra-thin van der Waals crystals as semiconductor quantum wells. *Nat. Commun.* **11**, 125 (2020).

36. Wang, Z. *et al.* Origin and Magnitude of `Designer' Spin-Orbit Interaction in Graphene on Semiconducting Transition Metal Dichalcogenides. *Phys. Rev. X* **6**, 041020 (2016).

37. Yang, B. *et al.* Tunable spin–orbit coupling and symmetry-protected edge states in graphene/WS2. *2D Mater.* **3**, 031012 (2016).

38. Patterson, C. L. *et al.* Superconductivity and spin canting in spin–orbit-coupled trilayer graphene. *Nature* **641**, 632–638 (2025).

39. Auerbach, N. *et al.* Isospin magnetic texture and intervalley exchange interaction in rhombohedral tetralayer graphene. *Nat. Phys.* **21**, 1765–1772 (2025).

40. Garreis, R. *et al.* Long-lived valley states in bilayer graphene quantum dots. *Nat. Phys.* **20**, 428–434 (2024).